\documentclass[12pt]{article}
\pdfoutput=1
\usepackage{cite}
\usepackage{amsmath,subfigure}
\usepackage{amssymb}
\usepackage{epsfig}
\usepackage{float}
\usepackage{color}
\usepackage{latexsym,graphics}
\usepackage{dsfont}

\def\sqr#1#2{{\vcenter{\vbox{\hrule height.#2pt\hbox{\vrule width.#2pt height#1pt \kern#1pt\vrule width.#2pt}\hrule height.#2pt}}}}

\title{Clonal selection and therapy resistance in acute leukemias: Mathematical modelling explains different proliferation patterns at diagnosis and relapse}
\author{Thomas Stiehl{$^1$}, Natalia Baran{$^2$}, Anthony D. Ho{$^2$}\\ and Anna Marciniak-Czochra{$^1$},\\
{\footnotesize $^1$ \it{Institute of Applied Mathematics, BIOQUANT and IWR}},\\ 
{\footnotesize \it{University of  Heidelberg, Im Neuenheimer Feld 294, 69120  Heidelberg, Germany}}\\
{\footnotesize $^2$\it{Department of Medicine V, University of Heidelberg}},\\
{\footnotesize \it{ Im Neuenheimer Feld 410, 69120 Heidelberg, Germany}}\\
\\
Published in J R Soc Interface 11(94):20140079, 2014.\\ doi: 10.1098/rsif.2014.0079. }
\date{}

\begin{document}
\maketitle
\abstract
{Recent experimental evidence suggests that acute myeloid leukemias may originate from multiple clones of malignant cells. Nevertheless it is not known how the observed clones may differ with respect to cell properties such as proliferation and self-renewal. There are scarcely any data on how these cell properties change due to chemotherapy and relapse.   We propose a new mathematical model to investigate the impact of cell properties on multi-clonal composition of leukemias. Model results imply that enhanced self-renewal may be a key mechanism in the clonal selection process. Simulations suggest that fast proliferating and highly self-renewing cells dominate at primary diagnosis while relapse following therapy-induced remission is triggered mostly by highly self-renewing but slowly proliferating cells. Comparison of simulation results to patient data demonstrates that the proposed model is consistent with clinically observed dynamics based on a clonal selection process. }\\

\section{Introduction}
Leukemia is a clonal disease of the hematopoietic system leading to extensive expansion of malignant cells that are non functional and cause impairment of blood cell formation. Recent experimental evidence indicates  that the malignant cell population might be composed of multiple clones \cite{Ding}, maintained by cells with stem-like properties \cite{Bonnet, Hope}.  A clone consists of genetically identical stem and non-stem cells. Relapse of the disease after therapy is a common problem of   leukemias \cite{Ding}. 

To understand better origins of acute leukemia relapses,  a genetic interdependence between clones at diagnosis and relapse has been investigated using gene sequencing and other techniques.  In most cases of acute lymphoblastic leukemia (ALL) the clones dominating relapse have already been present at diagnosis but undetectable by routine methods  \cite{VanDelft, Choi, LutzLeuk}.  Due to quiescence, very slow cycling or other intrinsic mechanisms,  \cite{LutzLeuk, Choi} these clones survive chemotherapy and eventually expand  \cite{LutzLeuk, Choi}. This implies that the main mechanism of relapse in ALL is based on a selection of existing clones and not an acquisition of therapy-specific mutations \cite{Choi}. Similar mechanisms have been described for acute myeloid leukemia (AML), where clones at relapse are genetically closely related to clones at primary diagnosis \cite{Ding, Jan} and did not have to acquire additional mutations during the course of disease \cite{Parkin, Bachas}.

  Based on these findings the evolution of malignant neoplasms can be interpreted as a selection process \cite{Sprouffske, Podlaha, Greaves} of cells with properties that enable them to survive treatment and to expand efficiently. Cells with different mutations may have different growth properties \cite{Ding}. Chemotherapy significantly alters growth conditions of cells and therefore, it may  have  a strong impact on the selection process. If cells dominating at diagnosis are sensitive to therapy,  minor clones with intrinsic resistance \cite{LutzLeuk, Choi, Klumper} may expand more efficiently once the competing clones are eliminated by the treatment.  The latter could explain manifestation of different cell clones at diagnosis and at relapse without a need for additional mutations in between. 
  
The mechanism of the underlying selection process and its impacts on the disease dynamics and on the response of cancer cells to chemotherapy are not understood.  Gene sequencing studies allow to decipher the genetic relation between different clones, nevertheless the impact of many detected mutations on cell behaviour remains unclear \cite{Ding} and often passenger mutations cannot be distinguished from relevant genetic changes \cite{VanDelft}. Many authors, e.g. \cite{Ran, Anderson}, have provided evidence for the heterogeneity of LSCs attempting to identify the LSC characteristics, for review see \cite{Lutz}. This heterogeneity is further supported by the results of gene sequencing studies \cite{Ding, Mardis, Ley}.
The multifactorial nature of the underlying processes severely limits the intuitive interpretation of experimental data.  Mathematical modelling is a powerful technique to close this gap and to provide quantitative insights into cell kinetics, fate determination and development of cell populations. It allows a systematic study of  processes not yet accessible by experimental procedures.
Mathematical models have been widely applied to analyze the regulatory mechanisms controlling the hematopoietic system and its diseases: for review  see   \cite{Roeder4, Mackey2, Whichard, Manesso}  and references therein.

The aim of this work  is to investigate  the impact of cell growth properties on the clonal selection process in acute leukemias before and after treatment. We introduce mathematical models of dynamics of leukemia, which are extended versions of the models proposed earlier by our group \cite{MarciniakStiehl, StiehlMarciniak, StiehlMarciniak2}.  The novel ingredients of the models in this work are: {(i) heterogeneity and multiclonal structure of LSCs,  (ii) different plausible feedback mechanisms, and (iii) effects of chemotherapy.}

 Since the mechanism of  interaction between healthy and leukemic cell lines is not well identified, we propose two models (see Fig.\ref{Fig1}). In the first one, we assume that leukemic cells depend on hematopoietic growth factors and interact with hematopoietic cells via competition for these factors. The second model is based on the assumption that autonomous leukemic clones compete with hematopoietic cells for niches in bone marrow, which leads to an increased cell death due to over-crowding.  The latter is supported by experimental findings showing signal-independent activation of important cell functions \cite{Hayakawa, Reilly, Polak} and by an increased cell degradation observed in leukemic patients \cite{Fanin, Buechner, Kornberg}. Such interactions have not been considered in previous models.

   The models proposed in this paper do not account for new mutations. Motivated by the experimental findings described above \cite{LutzLeuk, Choi,Parkin, Bachas}, we rather aim to understand which aspects of the dynamics of leukemias can be 
 explained by a selection process alone. It is interesting, since expansion of a clone at relapse that could not be detected at diagnosis due to limited sensitivity of methods can be misinterpreted as occurrence of mutations \cite{Choi}. This scenario seems to be relevant in case of acute leukemias with a  short duration treatment administration. Many acute leukemias are genetically relatively stable in comparison to other cancers \cite{Welch, JanLeuk}.  For this reason on average many replications are necessary to acquire a new mutation.  Consequently,  it is less probable that cells acquire mutations during short treatment and, therefore, intrinsic resistance to therapy may be important, as suggested by available evidence \cite{Choi}. The latter does not hold true for long-term drug administration, such as imatinib treatment in case of chronic leukemias.  
 For this reason,  our work focuses on the acute leukemias. 
 For completeness of this work and to check how mutations might influence the model dynamics as it concerns results presented in this paper,  we have developed a version of the model with mutations. The simulations of the model confirm our conclusions for the model with mutations. The model and simulations are presented in Appendix.

Using mathematical models, we aim to identify which cell properties are compatible with intrinsic resistance to therapy and efficient expansion after treatment, and to compare them with the cell properties selected for before treatment.  We perform computer simulations describing evolution of multi-clonal population of leukemic cells during the disease development and the contribution of different clones to the entire cancer cell population at different time points. The heterogeneity of the system is given by a certain number of leukemic clones already present at the beginning of our observation. The models provide information on the influence of cell properties on the growth dynamics of the different clones in presence or in absence of chemotherapy. This allows to understand how the cell properties selected for before treatment differ from those selected for during and after treatment and how treatment could be optimised to reduce relapses. Finally, we compare qualitatively model simulations to patients' data from clinical routine to show that the proposed models are consistent with clinical observations concerning the response to therapy and the time intervals between relapses of the disease. Details of mathematical formulation and parametrisation of the proposed models are presented in Appendix.

\section{Methods}
\subsection{Mathematical Models}
\subsubsection{Model assumptions}
The models used in this study are  based on the models 
of healthy hematopoiesis proposed and analysed in \cite{MarciniakStiehl, MarciniakStiehl2,Nakata,Getto2} and extended to account for evolution of a single leukemic clone in  \cite{StiehlMarciniak2}.

Based on the classical understanding of hematopoiesis \cite{Jandl}, we assume that the system consists of an ordered sequence of different maturation states, so called compartments.  To describe time evolution of cell populations, we apply ordinary differential equations. The enormous amount of cells forming the hematopoietic system justifies this approach  \cite{Jandl, Lansdorp}. Evolution of small cell population in the post-therapy period is modeled by cutting off the initial data which are below a minimal threshold, as it was, for example,  proposed in \cite{Brenes}.

We model time dynamics of one healthy cell lineage and an arbitrary number of leukemic clones. In the description of cell differentiation within each cell line, we choose a two-compartment version of the multi-compartment system established in \cite{MarciniakStiehl}. The model focuses on the maintenance of primitive cells and  differentiation from undifferentiated, proliferating  cells to differentiated, post-mitotic cells. In the case of healthy hematopoiesis, the proliferating cells are  hematopoietic stem cells (HSC), hematopoietic progenitor cells (HPC) and precursor cells, the post-mitotic cells are mature cells, e.g., white cells. The two-compartment architecture is based on a simplified description of the multi-stages differentiation process. Nevertheless as shown in \cite{StiehlMarciniak}, \cite{Nakata}, and \cite{Getto2} models consisting of two compartments capture the desired dynamics of the multi-compartmental cell population. This allows to reduce the complexity of the differentiation process to focus on mechanisms and effects of competition between different cell lines.

Each proliferating cell type is characterised by the following cell properties:  
\begin{itemize}
\item Proliferation rate, describing how often a cell divides per unit of time. 
\item  Fraction of self-renewal, describing the fraction of daughter cells returning to the compartment occupied by the mother cells that gave rise to them. Based on our earlier work and on compatibility with clinical data \cite{MarciniakStiehl}, we assume that the fraction of self-renewal of hematopoietic cells is regulated by feedback-signalling. 
\item Death rate, describing what fraction of cells dies per unit of time. For simplicity, we assume that under healthy conditions proliferating cells do not die and post-mitotic mature blood cells  die at a constant rate. We assume the same for leukemic cells in Model 1, while in Model 2 (see below), we consider, additionally,
cell density- dependent death rates for all bone marrow cell types if the marrow space is overcrowded. The considered marrow cell types include immature hematopoietic cells and mitotic and post-mitotic leukemic cells.  Overcrowding is defined when marrow cell counts exceed the steady state marrow cell count for 2 to 3 times.   In this case the death rate of post-mitotic leukemic cells consists of their intrinsic death rate and the death triggered by spatial competition.
\end{itemize}

Production of healthy blood cells is regulated by a negative feedback \cite{Layton, Metcalf, Fried}, mediated by cytokines,  such as G-CSF or EPO \cite{Fried, Jandl, Aglietta}. If there is a shortage of blood cells of a certain type, the concentration of signalling molecules increases and stimulates expansion of precursor cells. This effect is modeled using a negative feedback loop as proposed in \cite{MarciniakStiehl}. Analysis and simulation of the model of healthy hematopoiesis, validated based on the clinical observations after stem cell transplantations \cite{MarciniakStiehl, MarciniakStiehlProc, StiehlMarciniakBMT}, indicate that the regulation of the self-renewal is a more efficient mechanism than the regulation of the proliferation rates. Similar conclusions have been drawn using the models of multistage cell lineages applied to regeneration and maintenance of the mouse olfactory epithelium \cite{Lander1,Lander2}. Therefore, in the remainder of this paper we assume that the regulatory mechanism is based on the feedback inhibition of self-renewal depending on the level of mature cells.

\subsubsection{Model of the healthy cell line}
We denote by $p^c$  the proliferation rate of mitotic hematopoietic cells and by $a^c$ the corresponding fraction of self-renewal. The death rate of mature blood cells is denoted by $d_2^c$. We denote the concentration of healthy cell types  at time $t$ by $c_1(t), \ c_2(t)$, corresponding to mitotic and mature cells, respectively. The flux to mitosis  at time $t$ equals $p^c(t)c_1(t)$. During mitosis, a mother cell is replaced by two daughter cells. The outflux from mitosis at time $t$ equals $2p^c(t)c_1(t)$, of which  the fraction $2a^c(t)p^c(t)c_1(t)$ stays in compartment $1$ (process referred to as self-renewal). The fraction  $2\big(1-a^c_1(t)\big)p^c(t)c_1(t)$ moves to compartment $2$ (process referred to as differentiation). 

We denote the value of the feedback signal at time $t$ by $s(t)$, which takes values between zero and one. Self-renewal of a certain cell type at time $t$ is assumed to be given as a maximal possible self-renewal of this cell type multiplied by $s(t)$. Following \cite{StiehlMarciniak2, MarciniakStiehl}, we chose  $s(t)=\frac{1}{1+k^cc_2(t)}$, which can be derived from cytokine kinetics  \cite{ MarciniakStiehl}. The constant $k^c$ depends on the rate of extra-hematopoietic cytokine degradation by liver or kidney and on the rate of cytokine degradation by hematopoietic cells. The latter depends on the densities of cytokine receptors on hematopoietic cells \cite{StiehlMarciniakBMT}. 

We obtain the following system of ordinary differential equations, where $a^{c}_{max}$ corresponds to the maximal possible self-renewal of hematopoietic stem cells.
\begin{eqnarray}
\frac{d}{dt}c_1(t)&=&\big(2a^c_{max}s(t)-1\big)p^cc_1(t)\\
\frac{d}{dt}c_2(t)&=&2\big(1-a^c_{max}s(t)\big)p^cc_{1}(t)-d^c_2c_2(t)\\
s(t)&=&\frac{1}{1+k^cc_2(t)}
\end{eqnarray}
The two different models proposed in this manuscript differ with respect to the  interaction of leukemic and hematopoietic cells. We consider two cases. {In Model 1 leukemic cells depend fully on hematopoietic cytokines whereas in Model 2 they are totally independent of environmental signalling. In this sense Model 1 and Model 2 can be understood as the two opposite extremes of a continuum. In reality both mechanisms,  competition for environmental signals and direct inhibition or death of hematopoietic cells, may contribute to  impaired hematopoietic function \cite{Tsopra}. A schematic representation of the models is given in Fig \ref{Fig1}.

\subsubsection{Model 1}

We assume that leukemic cells depend on the same feedback signal as their healthy counterparts and that the post-mitotic leukemic cells (blasts) decrease the supply of the  factor. It describes a competition between healthy and leukemic cells for survival signals, which results in down-regulation of self-renewal.  A schematic representation of the model is given in Fig \ref{Fig1}.

To write the corresponding equations, we denote the number of leukemic clones by $n$. As for the hematopoietic cells we consider mitotic and post-mitotic cell compartments for each leukemic clone. Let $p^{l^i}$ denote the proliferation rate of mitotic cells in leukemic clone $i$ and $a^{l^i}_{max}$ the corresponding maximal  fraction of self-renewal. By $d_2^{l^i}>0$ we denote the clearance rate of post-mitotic cells of clone $i$. Denote by  $l_1^i(t)$ the level of mitotic cells of clone $i$ and by $l_2^i(t)$ the level of post-mitotic cells at time $t$.  These assumptions result in the following system of ordinary differential equations:
\begin{eqnarray}
\frac{d}{dt}c_1(t)&=&\big(2a^c_{max}s(t)-1\big)p^cc_1(t)\\
\frac{d}{dt}c_2(t)&=&2\big(1-a^c_{max}s(t)\big)p^cc_{1}(t)-d^c_2c_2(t)\\
\frac{d}{dt}l^1_1(t)&=&\big(2a^{l^1}_{max}s(t)-1\big)p^{l^1}l^1_1(t)\\
\frac{d}{dt}l^1_2(t)&=&2\big(1-a^{l^1}_{max}s(t)\big)p^{l^1}l^1_{1}(t)-d^{l^1}_2l^1_2(t)\\
\vdots&\vdots&\vdots\\
\frac{d}{dt}l^n_1(t)&=&\big(2a^{l^n}_{max}s(t)-1\big)p^{l^n}l^n_1(t)\\
\frac{d}{dt}l^n_2(t)&=&2\big(1-a^{l^n}_{max}s(t)\big)p^{l^n}l^n_{1}(t)-d^{l^n}_2l^n_2(t)\\
s(t)&=&\frac{1}{1+k^cc_2(t)+k^l\sum_{i=1}^n l^i_2(t)}.
\end{eqnarray}

{The expression for $s(t)$ is a special case of  $\tilde s(t)=1/\big(1+k^cc_2(t)+\sum_{i=1}^n k^l_il^i_2(t)\big)$, where we assume that $k^l_i=k^l$ for all $i$. This simplification corresponds to the observation that the density of cytokine receptors is similar on cells of all leukemic clones. For the major cytokine of the myeloid line, G-CSF \cite{Metcalf}, this is is true for many patients \cite{Shinjo}. Since there is evidence that  in some patients receptor densities may differ between different leukemic clones \cite{Shinjo}, we have repeated all simulations with a randomly chosen  $k^l_i$  value for each clone, ranging from 30\% of $k^c$  to 100\% of $k^c$. This heterogeneity had no significant impact on the model results. Since in many cases the receptor density on leukemic cells is of the same order of magnitude as that on hematopoietic cells \cite{Shinjo, Kondo}, we assume  also $k^l=k^c$ for the simulation of patient examples.

\subsubsection{Model 2}

{There is evidence that in some leukemias malignant cells show constitutive activation of certain signalling cascades and thus may become independent of external signals \cite{Hayakawa, Reilly, Polak}. We consider this scenario in Model 2.  In contrast to Model 1, we assume that leukemic cells are independent of hematopoietic cytokines, whereas the hematopoietic cell types depend on the nonlinear feedback described above.  Interaction between the healthy and cancer cell lines is modeled through a competition for space  resulting in an increased cellular degradation, for example due to overcrowded bone marrow space. This is consistent with the observation of an increase of markers for cell death such as LDH \cite{Fanin, Buechner, Kornberg}. Several mechanisms underlying this spatial competition have been proposed:
(i) physical stress due to overcrowding leads to extinction of cells (e.g., \cite{Griffin}; recently challenged by  \cite{Miraki}), (ii) competition for a limited niche surface expressing certain receptors (contact molecules) necessary for survival of the cells \cite{Calvi, Zhang} and apoptosis if no contacts to these molecules can be established \cite{Garrido}.

 We model the space competition by introducing a death rate that increases with the number of cells in bone marrow and acts on all cell types residing in bone marrow, i.e., mitotic and post-mitotic leukemic cells as well as mitotic hematopoietic cells. For simplicity we assume in Model 2 that all leukemic cells stay in bone marrow, since  the number of leukemic cells exiting bone marrow is highly variable among individuals and only partially dependent on the leukemia subtype \cite{Dommange, Berger2, Tavor2005} and since it is not well understood which mechanisms are responsible for marrow egress and high inter-individual variability.  The presented results are robust with respect to this assumption: We repeated all simulations for the cases that 10\%,  50\% or 90\% of the most mature leukemic blasts exit bone marrow. This has impact on the time dynamics of marrow blast count but does not influence the cell properties that are selected for.

Let $d(x)$ be an increasing function with $\lim_{x\rightarrow \infty}d(x)=\infty$. This function describes the death rates of bone marrow cells in dependence of bone marrow cell counts $x$. We assume that under healthy conditions there exists no cell death due to overcrowding. Enhanced cell death can be observed only if total bone marrow cellularity increases beyond the threshold level. This assumption is in line with bone marrow histology \cite{LoefflerAtlas}. Therefore, we assume that $d(x)=0$ for $x\leq\bar c_1$, where $\bar c_1$ is the steady state count of mitotic healthy cells.

Assuming that the hematopoietic cell lineage is regulated as described above,  we obtain the following system of differential equations:  
\begin{eqnarray}
\frac{d}{dt}c_1(t)&=&\big(2a^c_{max}s(t)-1\big)p^cc_1(t)-d(x(t))c_1(t)\\
\frac{d}{dt}c_2(t)&=&2\big(1-a^c_{max}s(t)\big)p^cc_{1}(t)-d^c_2c_2(t)\\
s(t)&=&\frac{1}{1+k^cc_2(t)}\\
\frac{d}{dt}\tilde{l}^1_1(t)&=&\big(2a^{\tilde{l}^1}-1\big)p^{\tilde{l}^1}\tilde{l}^1_1(t)-d(x(t))\tilde{l}^1_1(t)\\
\frac{d}{dt}\tilde{l}^1_2(t)&=&2\big(1-a^{\tilde{l}^1}\big)p^{\tilde{l}^1}\tilde{l}^1_{1}(t)-d^{\tilde{l}^1}_2\tilde{l}^1_2(t)-d(x(t))\tilde{l}^1_2(t)\\
\vdots&\vdots&\vdots\\
\frac{d}{dt}\tilde{l}^n_1(t)&=&\big(2a^{\tilde{l}^n}-1\big)p^{\tilde{l}^n}\tilde{l}^n_1(t)-d(x(t))\tilde{l}^n_1(t)\\
\frac{d}{dt}\tilde{l}^n_2(t)&=&2\big(1-a^{\tilde{l}^n}\big)p^{\tilde{l}^n}\tilde{l}^n_{1}(t)-d^{\tilde{l}^n}_2\tilde{l}^n_2(t)-d(x(t))\tilde{l}^n_2(t)\\
x(t)&=&c_1(t)+\sum_{i=1}^n \tilde{l}^i_1(t)+\sum_{i=1}^n \tilde{l}^i_2(t).
\end{eqnarray}
Here $\tilde{l}^i_1$ denotes the mitotic cells of clone $i$, $a^{\tilde{l}^i}$ their fraction of self-renewal and $p^{\tilde{l}^i}$ their proliferation rate. The level of post-mitotic cells of clone $i$ is denoted as  $\tilde{l}^i_2$. In absence of marrow overcrowding these cells die at rate $d^{\tilde{l}^i}_2$.    

\subsubsection{Chemotherapy}
We focus on classical cytotoxic therapy acting on fast dividing cells, which is introduced to the models by adding a death rate proportional to the proliferation rate. The assumption is motivated by the fact that many of the classical therapeutic agents used for treatment of leukemias act on cells in the phase of division or DNA replication, \cite{Berger}. Therefore, the rate of induced cell death is proportional to the number of cycling cells. We assume that the linear factor, denoted by $k_{chemo}$, is identical for all mitotic cells.
Under chemotherapy, the equation for mitotic hematopoietic cells in Model 1 takes the form
\begin{equation}\frac{d}{dt}c_1(t)=\big(2a^c_{max}s(t)-1\big)p^cc_1(t)-k_{chemo} \cdot p^c \cdot c_1(t).\end{equation} Similarly,  we obtain for mitotic cells of leukemic clone $i$
\begin{equation}\frac{d}{dt}l^i_1(t)=\big(2a^{l^i}_{max}s(t)-1\big)p^{l^i}l^i_1(t)-k_{chemo} \cdot p^{l^i}\cdot l^i_1(t).\end{equation}
Chemotherapy in Model 2 is introduced analogously.

\subsection{Simulations}
We perform numerical simulations of the models  to  investigate which leukemic cell properties lead to survival advantage during evolution of leukemogenesis and recurrence under chemotherapy.  As explained before, the models do not account for additional mutations taking place during the therapy. Instead, we investigate evolution of  a certain number of leukemic clones present at a starting time point. We assume that in healthy individuals the hematopoietic cells are in a dynamic equilibrium, i.e. production of each cell type equals its clearance. Initial conditions for the computer simulations are equilibrium cell counts in the hematopoietic cell lineage and a small cell number for different leukemic clones. We assume that the initial number of leukemic clones in each patient is 50. This number is arbitrarily chosen. All presented simulations were repeated for different numbers of leukemic clones (between 3 and 100), what led to comparable results (see Supplemental Figures \ref{SFig1}  and \ref{SFig2}). We assume that primary diagnosis and diagnosis of relapse occur, when healthy blood cell counts are decreased by 50\% of their steady state value. We perform simulations for 50 patients, i.e. 50 different sets of initial data and model parameters, with 50 leukemic clones per patient.  The growth properties of the leukemic clones are chosen randomly within certain ranges. The choice of model parameters is described in the Appendix. The simulations follow the following algorithm: 
\begin{itemize}
\item[(i)] We start from healthy equilibrium in the hematopoietic lineage and one mitotic cell per kg of body weight for each leukemic clone and run simulations until the number of healthy mature blood cells decreases by 50\%. We investigate properties of the clones with the highest contribution to the total leukemic cell mass.  The clones under consideration are those which together constitute  80\% of the total leukemic cell mass.  In the following we denote these clones as '{\it significantly contributing clones}'.  This procedure is taken to reflect the sensitivity of the detection methods. In more than 90\% of the patients 2 to 4 clones sum up to more than 95\% of the total leukemic cell mass. Taking a threshold between 80\% and 95\% to define '{\it significantly contributing clones}'  has little influence on the result. Furthermore, more than 97\% of the clones that are considered as insignificant by this method consist of less than 1\% of the leukemic cell mass. This number is in agreement with the detection efficiency reported in literature \cite{Meyer}.
\item[(ii)] Next, we simulate chemotherapy. For simplification, we consider seven applications of cytotoxic drugs (one per day during seven following 
days, corresponding to standard inductions).  Simulations show that the number of drug applications has no influence on the presented qualitative results.  As proposed in literature \cite{Brenes}, we assume that a cell population has become extinct if it consists of less than one cell.  Initial conditions for the post-therapy period are obtained from cell counts after therapy where counts of extinct populations are set to zero. We continue simulations until mature blood cell counts decrease by 50\% and then assess the cell properties of the clones contributing to relapse.
\end{itemize}

Calibration of the hematopoietic part of the model to clinical data and parameters for simulation of two patient examples can be found in Appendix.  Since in clinical routine only few key mutations are monitored, we choose patient examples with different key mutations detected at diagnosis and at relapses. Such data is relatively rare, therefore we focus on two patients.  Simulations are performed using standard ODE-solvers from the Matlab-software package (Version 7.8, The MathWorks, Inc, Natic, MA) which are based on Runge-Kutta schemes.

\section{Results}
\subsection{Clonality at diagnosis}
We solve the models numerically to obtain insight into the contribution of different leukemic clones to the total leukemic cell mass. Simulations indicate that at the diagnosis rarely more than 3-4 clones significantly contribute to the total leukemic cell mass. In most cases more than 40-50\% of the total leukemic cell mass originates from a single leukemic clone. This finding is identical for both considered models.
\subsection{Properties of clones at diagnosis}
Simulations indicate that the clones significantly contributing to the leukemic cell mass have high proliferation rates and high self-renewal potential (high fraction of symmetric self-renewing divisions). Such configuration of parameters leads to an efficient cell expansion. The properties of clones contributing significantly to leukemic cell mass at diagnosis are depicted in Figure \ref{Fig2}. This finding is identical for both considered models.

\subsection{Clonality at relapse}
The clonality at relapse is comparable to the clonality at diagnosis. Rarely more than three clones significantly contribute to the total leukemic cell mass. This finding is the same for both considered models.
\subsection{Properties of clones at relapse}
The properties of the leukemic clones responsible for relapse depend on the efficiency of chemotherapy. We run computer simulations for varied efficiency of chemotherapy, namely different death rates imposed on mitotic cell compartments. In the case of inefficient chemotherapy, i.e. killing rates of mitotic cells being relatively small, the clones present at diagnosis are also responsible for relapse. These clones have high proliferation rates and high self-renewal potential. In the case of more efficient chemotherapy, i.e. killing rates of mitotic cells being higher, the clones responsible for primary presentation differ from the clones responsible for relapse. Compared to the clones leading to primary presentation, the clones responsible for relapse have low proliferation rates but high self-renewal potential. The properties of clones contributing significantly to leukemic cell mass at diagnosis and at relapse are depicted in Figure \ref{Fig3}. Both models lead to similar results.

The result that slow cycling is an important selective  mechanism is compatible with the finding that cells in minimal residual disease samples are highly quiescent \cite{LutzLeuk}. It is further supported by the fact that addition of anthracyclines, which act independent of cell cycle \cite{Dy},  leads to improved outcome of relapse therapies in  ALL \cite{Chessells}.

\subsection{Treatment of relapse}

If the same treatment strategy as in case of primary treatment is applied to a relapsed patient, remission time is significantly shorter (Fig. \ref{Fig4}). Second relapse is mostly triggered by the same clones as primary relapse. With repeated chemotherapy, clonal composition changes in favor of the clones with minimal proliferation  (Clone 5 in Fig. \ref{Fig4}). 
This finding is in agreement with data from clinical practice in ALL suggesting  that the clones selected for at relapse possess inherently reduced sensitivity to treatment \cite{Choi} and may be also responsible for second relapse \cite{Choi}.   The dynamics of leukemic cells in our model are in good agreement with data from clinical practice: Chemotherapy is able to reduce leukemic cell load after relapses \cite{VanDelft}, nevertheless this reduction does not lead to durable remission \cite{Chessells}. This reflects the worse prognosis of relapsed patients \cite{Klumper,  Bhatla, Chessells}. The increasing fraction of cells with reduced drug sensitivity predicted by the simulations explains the experimental finding that cells present at relapse are more resistant to chemotherapy than cells present at initial diagnosis \cite{Klumper, Bhatla}. It also shows that repetition of the same induction therapy leads to worse results in relapse compared to primary manifestation \cite{Chessells}.  The selection of slowly cycling cells predicted by our model seems to be an important mechanism in AML. It it was demonstrated that induction of cell cycling enhances chemo-sensitivity of leukemic cells  \cite{Saito} and improves patient outcome after therapy \cite{Loewenberg}.  Our model suggests that repeated chemotherapy can lead to the selection of clones that are not competitive in natural environment, i.e. that can be outcompeted by clones sensitive to chemotherapy after cessation of the treatment.

\subsection{Short term expansion efficiency does not correlate with long term self-maintenance}
If leukemic cell behaviour depends on hematopoietic cytokines (Model 1), the current signalling environment influences expansion of leukemic clones. In this scenario it is possible that fast proliferating cells with low self-renewal potential dominate the leukemic cell mass during an initial phase. If, with increasing leukemic cell mass, self-renewal becomes down-regulated, e.g. due to occupation of bone marrow niche, eventually the cell clone with the highest affinity to self-renewal survives, although its proliferation might be slow. An example of time evolution during  an early phase is depicted in Fig. \ref{Fig5}.

\subsection{Late relapses can originate from clones that were already present at diagnosis}
Simulations of Model 1 indicate that late relapses, e.g.,  relapses after more that 3 years, can originate from clones that were already present at diagnosis but did not significantly contribute to the leukemic cell mass at that time. These relapses are triggered by very slow proliferating cells which survive chemotherapy and then slowly grow. At primary diagnosis fast proliferating clones dominate. The slowly proliferating clones are then selected by chemotherapy. This finding is able to explain relapses without additional mutations occurring after primary diagnosis. Thus, temporary  risk factor exposure (e.g., chemicals or radiation) can be responsible also for very late relapses and presentations. 

\subsection{Comparison of simulations to patient data}
To check if  the proposed modelling framework is consistent with the observed dynamics of leukemia, we calibrate the model to data of two patients with multiple relapses. The selected two patients showed different AML-typical mutations. Properties of leukemic cells and their impairment due to chemotherapy cannot be measured directly and the effects of specific mutations on cell dynamics are not well understood. {The available data include time periods between induction/consolidation chemotherapy and relapse as well as the percentages of leukemic blasts in the bone marrow at diagnosis, follow-ups and relapse. In addition emergence and subsequent elimination of leukemia driving mutations (FLT3, MLL-PTD) in the bone marrow cells were precisely monitored using molecular biology methods \cite{Schnittger, Thiede, Weisser}.  We verify if, and under which assumptions concerning the cell behaviour, the proposed model is compatible with clinical observations.} This can serve as a qualitative 'proof of principle' and leads to hypotheses concerning changes in cell properties induced by the respective mutations. We assume that each mutation is associated with one leukemic cell clone. We interpret differences at diagnosis and at relapse as the result of a clonal selection process due to chemotherapy and cell properties. For this study we apply Model 2, since simulations over a large range of parameters showed that remissions shorter than 150 days are only compatible with Model 2.

Simulations of the evolution of leukemic clones in the two patients are depicted in Figures \ref{Fig6} and \ref{Fig7}. The results show that bone marrow blast fraction can be well described by the model. In Patient 1  FLT3-ITD mutation of a length of 39 bp is detected at diagnosis. This mutation becomes extinct and the relapse is triggered by  two different  FLT3-ITD mutations (42 bp and 63 bp). This behaviour is reproduced in the model simulation. At diagnosis leukemic cell mass is mainly derived from one clone while at relapse two different clones contribute to leukemic cell mass.

In Patient 2 FLT3-ITD mutation and MLL-PTD-mutation were both detected at diagnosis. The MLL-PTD mutation practically did not contribute to relapse. The model reflects this scenario.  At diagnosis two different clones contribute to leukemic cell mass, one of which becomes extinct and is not detected at the relapse. In this patient the clone responsible for relapse behaves similarly to the HSC lineage.  Thus, classical cytotoxic treatment would not lead to its eradication. This is an indication for application of new anti-leukemic drugs, if feasible, or for bone marrow transplantation.

\section{Discussion}
We have examined the impact of cell properties on clonal evolution in acute leukemias during the course of disease. We have considered two different mathematical models, representing different modes of interactions between normal hematopoietic and leukemic cells. In Model 1 leukemic cells depend on hematopoietic cytokines, niches or other environmental factors. In Model 2 the leukemic cells are independent of these aforementioned determinants and the only interaction between benign and malignant cells is due to a competition for bone marrow space. 

Model simulations suggest that clones with a high proliferation rate and a high self-renewal are favoured at primary diagnosis.  The results indicate that the number of clones significantly contributing to the leukemic cell mass is relatively small, even if a large number of clones with different leukemia driving mutations might coexist in the bone marrow.  For example, in our simulations it was reduced from 50 to 2-5. This result is in agreement with data from recent gene sequencing studies and allows to explain these data. In these studies  \cite{Ding, Anderson} at most 4 contributing clones were detected in case of AML and at most 10 in case of ALL. In many patients this number was even smaller.     Our study implies that clonal selection due to different growth characteristics is an efficient mechanism to reduce the number of clones contributing to leukemic cell burden. Clones not  contributing to primary disease manifestations might rest in a slowly proliferating or quiescent state and expand at relapse. Chemotherapy exerts a strong selective pressure on leukemic clones and thus has a considerable impact on the clonal composition during relapse.

In the case of insufficient chemotherapy, the relapse can be triggered by the same clones as the primary disease. In the case of more intensive therapy regimens, relapses are mostly triggered by different clones than primary disease.  This has also been concluded from experimental studies \cite{Ding}. Our models suggest that chemotherapy selects for slowly proliferating clones with high self-renewal property. Depending on efficiency of the therapy, it is also possible that clones with high proliferation and high self-renewal potential are responsible for relapse.

 In the present study we have focused on classical cytotoxic chemotherapy, mostly acting on mitotic cells. This explains selection of slowly proliferating clones, among which those with high self-renewal potential have a competitive advantage, as shown in earlier studies \cite{MarciniakStiehl, StiehlMarciniak, StiehlMarciniak2}. High proliferation rates  constitute a disadvantageous factor under cytotoxic treatment, since fast proliferating cells are responsive to even moderately intensive therapy regimens. Relapses due to such clones are only possible if LSCs at  the same time have a high self-renewal potential, which is an advantageous factor for expansion and survival. Otherwise they would be out-competed by slowly proliferating cells with high self-renewal. Fast proliferating cells with low self-renewal have never been observed at relapse in our simulations. Their emergence at relapse could only be explained by additional mutations acquired after initial treatment. The selection of slowly proliferating cells may explain emergence of resistance in relapses. In such case, applying an identical therapeutic regimen to primary presentation and relapse has limited effects in the absence of new mutations.
  
   The principle of clonal competition in leukemia evolution and the fact that resistant subclones might be responsible for relapse have been discussed for a long time \cite{Choi}. Using mathematical modelling, we have provided for the first time evidence that self-renewal potential is a major force behind this mechanism and that cells responsible for a relapse show high self-renewal in nearly all cases. This finding is new and cannot be concluded from biological data so far.\\

   In the Appendix we study a model that includes occurrence of new mutations in addition to the selection process. In this scenario the number of clones detectable at diagnosis and at relapse and their respective properties are practically identical to the scenario without mutations. This finding underlines that clonal selection has an important impact on the evolution of leukemic cell properties. 

The exact nature of interaction between leukemic and hematopoietic cells is not well understood. Moreover, it is well known that leukemias show high inter-individual heterogeneity concerning symptoms and survival \cite{Liesveld}. Therefore, it is possible that different mechanisms may be relevant in different cases. Simulation results suggest that the evolving cell properties  are robust with respect to the assumptions on the exact mode of interaction between hematopoietic and leukemic cells and are similar in different scenarios and different patients.
 Common features of both models are: (i) Relapses can be explained by cells that were already present at diagnosis. (ii) Before therapy clonal evolution selects for cells with high proliferation rate and high self-renewal.  (iii) Cytotoxic treatment selects for cells with slow proliferation and high self-renewal. Thus, it is possible to draw conclusions on leukemic cell properties, even if their interaction with the healthy hematopoiesis is not known in detail.  
  
 Nevertheless the two proposed models exhibit some different dynamical properties, namely: (i) Complete remissions lasting shorter than 150 days  are only possible in Model 2. (ii) In Model 2 it is possible that leukemic and non-leukemic cells coexist at ratios compatible with sufficient hematopoiesis for long times.  (iii) In Model 1 clones can temporarily expand and then be outcompeted. In Model 2 the clone with fastest expansion is dominant for all times until treatment. (iv) In Model 2 leukemic cell load can be reduced to a new steady state under chronic application of cytostatic drugs. In Model 1 expansion of leukemic cells can be reduced in speed but eventually healthy hematopoiesis will be outcompeted. This may have application in treatment of fast relapsing patients, since fast relapse can only be explained by Model 2.
 
Up to now it cannot be decided which model is more realistic. For each of the models there exist supportive findings. Model 1 is supported by observations on expression of growth factor receptors by leukemic cells similar to those by hematopoietic cells \cite{Shinjo, Kondo}, expansion of leukemic cells in presence of cytokines in some patients \cite{Vellenga} and dependence of leukemic cell self-renewal and proliferation on  chemokines needed for hematopoietic cell maintenance  \cite{Tavor2008}. The facts supporting Model 2 are enhanced cell death in marrow samples \cite{Irvine} and increased markers for cell death/cell lysis in serum \cite{Fanin, Yamauchi}, independence of leukemic cells from  important environmental signalling cues in presence of some mutations \cite{Hayakawa} and necessity of  physical contacts  to marrow stromal cells needed for cell survival  \cite{Calvi, Zhang, Garrido}).

Our models support the hypothesis that  processes of clonal selection are important mechanisms of leukemia relapse, which can be responsible for expansion of different cell clones without a need for new mutations. A testable prediction of our models is that more sensitive methods should reveal larger numbers of different clones that exist but do not significantly contribute to the leukemic cell mass.  Another prediction is that cells present at relapse show mutations responsible for high self-renewal.
 
Calibration of the models to patient data shows that the proposed framework is compatible with the observed clinical course in the considered two data sets. The predicted selection of slowly proliferating cells with high self-renewal ability is consistent with clinical observations. Our results may have relevance for personalised medicine. Deep sequencing techniques might provide information on the genetic interdependence of the clones present at diagnosis and relapse \cite{Ding}. Our model suggests that insufficient therapy may lead to presence of the same clones at diagnosis and relapse. If the clones present at diagnosis and relapse are not identical but related, i.e., they share common somatic mutations \cite{Ding}, relapse may be due to a selection process. In this case it is probable that the clones present at relapse show a slow proliferation and a high self-renewal.  One possible implication might be the application of cell-cycle independent drugs, such as those used in targeted therapies.

\section*{Acknowledgements}
This work was supported by the Collaborative Research Center, SFB 873 "Maintenance and Differentiation
of Stem Cells in Development and Disease". AM-C was supported by ERC Starting Grant
”Biostruct” and Emmy-Noether-Programme of German Research Council (DFG).  The authors 
would like to thank Professor Marek Kimmel for many helpful advice during 
preparation of the manuscript.

\section{Appendix}

\subsection{Calibration of the hematopoietic cell lineage}

In absence of leukemic clones, both considered models reduce to the same model of the hematopoietic system. In steady state this  model has the following form
\begin{eqnarray}
0&=&(2a^c\bar{s}-1)p^c\bar{c}_1\\
0&=&2(1-a^c\bar{s})p^c\bar{c}_1-d^c_2\bar{c}_2\\
\bar{s}&=&\frac{1}{1+k^c\bar{c}_2}
\end{eqnarray}
Assume we know $\bar{c}_1$ and $\bar{c}_2$. 
It holds
\begin{eqnarray}
\frac{\bar{c}_2}{\bar{c}_1}&=&\frac{p^c}{d^c_2}\\
\bar{c}_2&=&\frac{2a^c-1}{k^c}
\end{eqnarray}
Knowing $a^c$, we can calculate $k^c={(2a^c-1)}/{\bar{c}_2}$, such that the steady state population size $\bar c_2$ is satisfied.
We calibrate the model to the data on production of neutrophil granulocytes, which constitute the majority of mature white blood cells ($50\%-70\%$). Lymphocytopoiesis is a complicated process involving lymphatic organs and not only the bone marrow \cite{Crooks}. Therefore, we restrict ourselves to the myeloid line. It holds for the steady state count of neutrophils \cite{Neumeister}, $\bar{c}_2 \in (3-5.8)\cdot 10^9/l$.  We interpret $\bar{c}_1$ as the total amount of mitotic neutrophil  precursors  in bone  marrow.  Based on the data from \cite{LoefflerAtlas} we assume that about 20\% of bone marrow cells are mitotic precursors of  neutrophils (the interindividual variations are considerable). The total bone marrow cellularity is about $10^{10}$ cells per kg of body weight \cite{Harrison}.
Therefore, we take \begin{equation}\bar{c}_1 \approx 2 \cdot 10^{9}/kg.\end{equation} Assuming an average blood volume of 6 liters and an average body weight of 70 kg, we calculate a mature neutrophil count of \begin{equation}\bar{c}_2 \approx 4\cdot 10^8/kg.\end{equation} Neutrophils have half-life in blood stream, $T_{1/2}$, of about 7 hours\cite{Cartwright}. From this we calculate \begin{equation}d^c_2=\frac{\ln(2)}{T_{1/2}}\approx 2.3/days.\end{equation}
We obtain
\begin{equation}p^c=\frac{\bar{c}_2}{\bar{c}_1}d^c_2\approx 0.45 / day,\end{equation}
i.e.,  about once per 1.5 days. We know from bone marrow transplantation \cite{Klaus} that patients need about 15 days to reconstitute to $5\cdot 10^8$ neutrophils per liter of blood  ($4\cdot 10^7$ per kg ) after infusion of $5 \cdot 10^6$ immature cells per kg of body weight.   For \begin{equation} a^c\approx 0.87\end{equation}
this constraint is met.

\subsection{Model parameters}
For the hematopoietic branch we chose parameters obtained from the calibration in the Section above. For simplicity we assume $k^c=k^l$ for the feedback mechanism in Model 1. We set the clearance rate of blasts (in absence of effects of overcrowding) to $d_2^{l^i}=0.1$. This is based on the apoptotic indices (fraction of dying cells) reported in literature which are $\approx 0.19 \pm 0.16$ $(19\% \pm 16\%)$ \cite{Malinowska, Savitskiy}. Choosing blast clearance between $0.1$ and $0.5$ changes the speed of leukemic cell accumulation but, as revealed by additional simulations, not the cell properties that are selected. 

We chose $d(x)=d_{const}\cdot \max\{0,x\!-\!x_{max}\}$. In histological images of healthy adult bone marrow a large part of the bone marrow cavity consists of fat and connective tissue and is free of hematopoietic cells. To reflect this fact, we set $x_{max} \approx 2\bar c_1$, where $\bar c_1$ is the steady state count of mitotic healthy cells. In the simulations, $d_{const}$ was set to $10^{-10}$. This choice implies that if bone marrow cell counts are three times higher than in the steady state, the additional death rate  due to overcrowding is of the order of magnitude of mature cell clearance. The results remain unchanged qualitatively, even if values of $d_{const}$ vary within different orders of magnitude.

We apply chemotherapy on seven following days for 2 hours. Different treatment intervals lead to comparable results.  In the depicted simulations $k_{chemo}$ has been set to values between $20$ and $30$ for less efficient therapy and to values between $40$ and $60$ for efficient chemotherapy. For different choices similar results are obtained. The higher $k_{chemo}$, the stronger the selection for high self-renewal and slow proliferation and the lower the probability of relapse. The lower $k_{chemo}$, the higher the probability that clones contributing to primary presentation are among clones contributing to relapse.  For $k_{chemo}$ between 40 and 45, the obtained results are similar to the results obtained in experiments \cite{Ding}. The parameters of the leukemic clones are chosen randomly from uniform distributions, assuming that cells divide at most twice per day and that self-renewal is between zero and one.
 
\subsection{Calibration to patient examples} \label{PatCalib}
Both patients were treated within clinical trials at the University Hospital of Heidelberg after obtaining their written consent. Details on the patients' characteristics and therapeutic regimens  can be found in Supplemental Table \ref{TabTherap}.  Model parameters can be found in Supplemental Tables \ref{TabPat1} and \ref{TabPat2}. For both patients the presence of specific key mutations was assessed in clinical routine.  We chose cases where mutations get lost due to treatment and new mutation are detected at relapse. We interpret this as  the result of clonal evolution. Since the two patients harbour different mutations, their leukemic cells can have different properties.   Chemotherapy is modeled by increasing death rates for mitotic cell types during the duration of each cycle. For simplicity, we did not model kinetics of single chemotherapeutic agents. Instead, the therapy-induced death rates are assumed to remain constant from the first to the last day of each treatment cycle. In pharmacology, the exposition to a drug is measured using the {\it 'area under the curve' (AUC)}. This is the integral of concentration (or drug effect) over time \cite{Goodman}. The AUC in our case is $k_{chemo}\cdot \Delta t$, where $\Delta t$ is the period of drug action. The AUC over one day of therapy is similar for the single patient examples and the simulations  in Figure \ref{Fig3}. Only myeloablative treatment before transplantation has a higher AUC. The presented results are based on Model 2. Model 1 is not compatible with remissions lasting less than 150 days.  For simplicity we count all leukemic cell types as blasts. 

\subsection{Model with mutations}

In the following we describe an extension of  Model 1 which includes mutations. There is an evidence that a preleukemic HSC compartment serves as a reservoir of accumulated mutations \cite{JanLeuk}. This hypothesis is supported by the finding that some of the mutations recurrently observed in leukemia already exist in the HSC compartment of a majority of leukemia patients \cite{JanLeuk}. These preleukemic HSC seem to behave similarly to normal HSC \cite{JanLeuk}. The hypothesis is that a relatively small number of additional hits may transform these preleukemic HSC into LSC. Nevertheless details of the underlying dynamics are not well understood \cite{JanLeuk}. 

We make the following assumptions:
\begin{itemize}
\item LSC with new properties can be generated either from preleukemic HSC or from LSC due to acquisition of mutations.  This is in line with the current knowledge \cite{Jan, Welch}. 
\item For simplicity we assume that the influx of new LSC from the preleukemic compartment is constant in time. This assumption is made due to simplicity, since at the moment the dynamics of the preleukemic compartment is not well understood \cite{JanLeuk}. We neglect mutations leading from normal HSC, i.e. non-preleukemic HSC to LSC. 
\item We assume that most mutations in LSC are acquired during replication of the genome and neglect other possible origins. In line with \cite{Jan} we neglect mutations leading to dedifferentiation of non-LSC leukemic cells.
\end{itemize}

Let $l_1^i(t)$ be the level of LSC of clone $i$ at time $t$. The flux to mitosis is then $l_1^i(t)p^{l^i}(t)$. Out of mitosis we obtain $2a^{l^i}(t)p^{l^i}(t)l_1^i(t)$, where $a^{l^i}(t)$ is the fraction of LSC self-renewal of clone $i$ at time $t$. We assume that the fraction $\nu$ of these cells is mutated, $\nu$ takes into account replication errors in relevant genes and is assumed to be constant. The influx $\alpha_i(t)$ of mutated LSCs  due to new mutations occurring in  clone $i$ at time $t$ is therefore    $2a^{l^i}(t)p^{l^i}(t)l_1^i(t)\nu$. 

We obtain the following set of equations describing dynamics of clone $i$:

\begin{align*}
\frac{d}{dt} l_1^i(t)&=2a^{l^i}(t)p^{l^i}(t)l_1^i(t)(1-\nu) -p^{l^i}(t)l^i_1\\
\frac{d}{dt}l_2^i(t)&=2(1-a^{l^i}(t))p^{l^i}(t)l_1^i(t)-d_2^{l^i}l_2^i(t)\\
\alpha_i(t)&=2a^{l^i}(t)p^{l^i}(t)l_1^i(t)\nu
\end{align*}
A similar system of equations has been obtained by Traulsen et al \cite{Traulsen}. Since $l^i_2$ is considered to be postmitotic, we do not distinguish between cells that acquired a mutation during the divisions and those that did not.

The influx $\alpha(t)$ of mutated cells at time $t$ is given by 

$\alpha(t)=\gamma+\sum_{i=1}^{N(t)} \alpha_i(t)$, where $\gamma$ is the constant influx from the preleukemic compartment and $N(t)$ the number of leukemic clones present at time $t$.

We accept the rate $\alpha(t)$ as the rate of an inhomogeneous Poisson process. Poisson processes describe rare events \cite{Ross, Ross2}, therefore they are a suitable tool to model mutations. It is known from probability theory that, if $\tau_1$ is a jump time of the inhomogeneous Poisson with rate $\lambda(t)$, then the next jump time $\tau_2$ can be generated by solving the equation $\int_{\tau_1}^{\tau_2}\lambda(t)dt=-\log (1-u)$ for $\tau_2$. Here $u$ is  an uniformly distributed random variable $u\in [0, 1]$ \cite{Klein}. We further know that  if $u$ is uniformly distributed in $u\in [0, 1]$ then $-\log (1-u)$  is exponentially distributed with parameter $1$ \cite{Ross2}.

We simulate the system with mutations as follows: At time $t_0=0$ we draw an exponentially distributed random number $r_1$ with parameter $1$. We simulate the system until the timepoint $t_1$ which fulfills $\int_{t_0}^{t_1}\lambda(t)dt=r_1$. At timepoint $t_1$ a mutation occurs that gives rise to a new LSC. This is modelled by adding to the system a new LSC clone, consisting of one LSC and no less primitive leukemic cells. We assume that the mutation occurs in a random gene position therefore we assign random cell properties to the new clone, i.e. self-renewal and proliferation chosen randomly from uniform distributions (proliferation rate between $0.01$ and $0.9$, self-renewal between $0.5$ and $1$).  This choice is made for the sake of simplicity since details about the impact of mutations on cell behaviour and the underlying probability distributions are not known \cite{Ding}. We then draw another random number $r_2$ and continue simulations until timepoint $t_2$ fulfilling $\int_{t_1}^{t_2}\lambda(t)dt=r_2$, etc. We start simulations from the equilibrium of the hematopoietic system and one LSC with random properties.

The results obtained from these simulations are similar to the results from the model without mutations. At primary diagnosis cells show high self-renewal and high proliferation while at relapse cells show high self-renewal and reduced proliferation (Supplemental Figure \ref{SFig3}). The proliferation rates differ significantly between diagnosis and relapse ($p<10^{-6}$ in Kruskal-Wallis Test), while self-renewal does not differ significantly ($p\approx 0.7$ in Kruskal-Wallis test).

{\footnotesize
\begin{table}[H]
\begin{tabular}{p{0.18\textwidth}|p{0.4\textwidth}p{0.42\textwidth}}
&\bf{\footnotesize Patient 1}&\bf{\footnotesize Patient 2}\\
\hline
\footnotesize Gender &Male&Male\\
\footnotesize Age at diagnosis &63&60\\
\footnotesize Diagnosis&AML &AML FAB M2\\
\footnotesize Therapy &Days 30-37 (Induction): \newline \footnotesize - Mitoxantron(10 $mg/m^2$) (d 1-3)\newline
\footnotesize- AraC (2000 $mg/m^2$) (d 1,3,5,7)\newline $ $\newline \footnotesize Days 70-74 (Consolidation): \newline\footnotesize  - Mitoxantron(10 $mg/m^2$) (d 1,2)\newline
\footnotesize - AraC (1000 $mg/m^2$) (d 1,3,5)\newline&\footnotesize Days 1-7 (Induction I)\newline\footnotesize  -AraC(100 $mg/m^2$)(d1-7)\newline \footnotesize -DA(60$mg/m^2$)(d3-5)\newline $ $\newline
\footnotesize Days 23-29 (Induction II)\newline \footnotesize -AraC(100 $mg/m^2$)(d1-7)\newline\footnotesize  -DA(60$mg/m^2$)(d3-5)\newline$ $\newline \footnotesize  Days 60-64 (Consolidation I)\newline \footnotesize -AraC(6000 $mg/m^2$)(d1,3,5)\newline$ $\newline \footnotesize Days 109-113 (Consolidation II) \newline \footnotesize -AraC(6000 $mg/m^2$)(d1,3,5)\newline$ $\newline \footnotesize Days 145-146 (Conditioning)\newline -\footnotesize Allogeneic, HLA-identical HSCT\newline \footnotesize (after Treosulfan/Fludarabin/ATG)
\end{tabular}
\caption{Demographic and treatment data of the 2 patients considered. Day 0 is defined as the day of diagnosis. ATG=Anti-Thymocyte Globulin, HSCT=Hematopoietic Stem Cell Transplantation.}\label{TabTherap}
\end{table}

\begin{table}[H]
\small
\begin{tabular}{l|lll}
{\bf Cell Property}& \bf{Clone 1}&\bf{Clone 2}&\bf{Clone 3}\\
\hline
Leukemic cell proliferation rate (1/days)&0.25 &0.25 &1.2 \\
Leukemic cell self-renewal &0.755 &0.76 &0.6\\
Blast death rate (1/days)& 0.5& 0.5&0.5\\
$k_{chemo}$ (Induction)&3 &3 &3\\
$k_{chemo}$ (Consolidation)&3 &3 &3\\
\end{tabular}
\caption{Parameters used for Patient 1: The parameters of the hematopoietic lineage result from the model calibration shown in the Supplement. Simulations are based on Model 2. The function $d(x)$, describing cell death due to space competition has been set to $\max(0,x-3c)$, where $c$ is steady state bone marrow cell count in absence of leukemic cells.}\label{TabPat1}
\end{table}

\begin{table}[H]
\begin{tabular}{l|ll}
\bf{Cell Property}&\bf{Clone 1}&\bf{Clone 2}\\
\hline
Leukemic cell proliferation rate (1/days)&0.45 &1.2\\
Leukemic cell self-renewal &0.75 &0.6\\
Blast death rate (1/days)& 0.5&0.5\\
$k_{chemo}$ (Induction I, II)& 3&3\\
$k_{chemo}$ (Consolidation I, II)&5 &5\\
$k_{chemo}$ (Conditioning)&16 &16
\end{tabular}
\caption{Parameters used for Patient 2: The parameters of the hematopoietic lineage are chosen due to the calibration shown in the Supplement. Simulations are based on Model 2. The function $d(x)$, describing cell death due to space competition has been set to $\max(0,x-3c)$, where $c$ is steady state bone marrow cell count in absence of leukemic cells.}\label{TabPat2}
\end{table}
}

\newpage

\begin{figure}[htpb] 
\begin{center} 
\includegraphics[width=\textwidth]{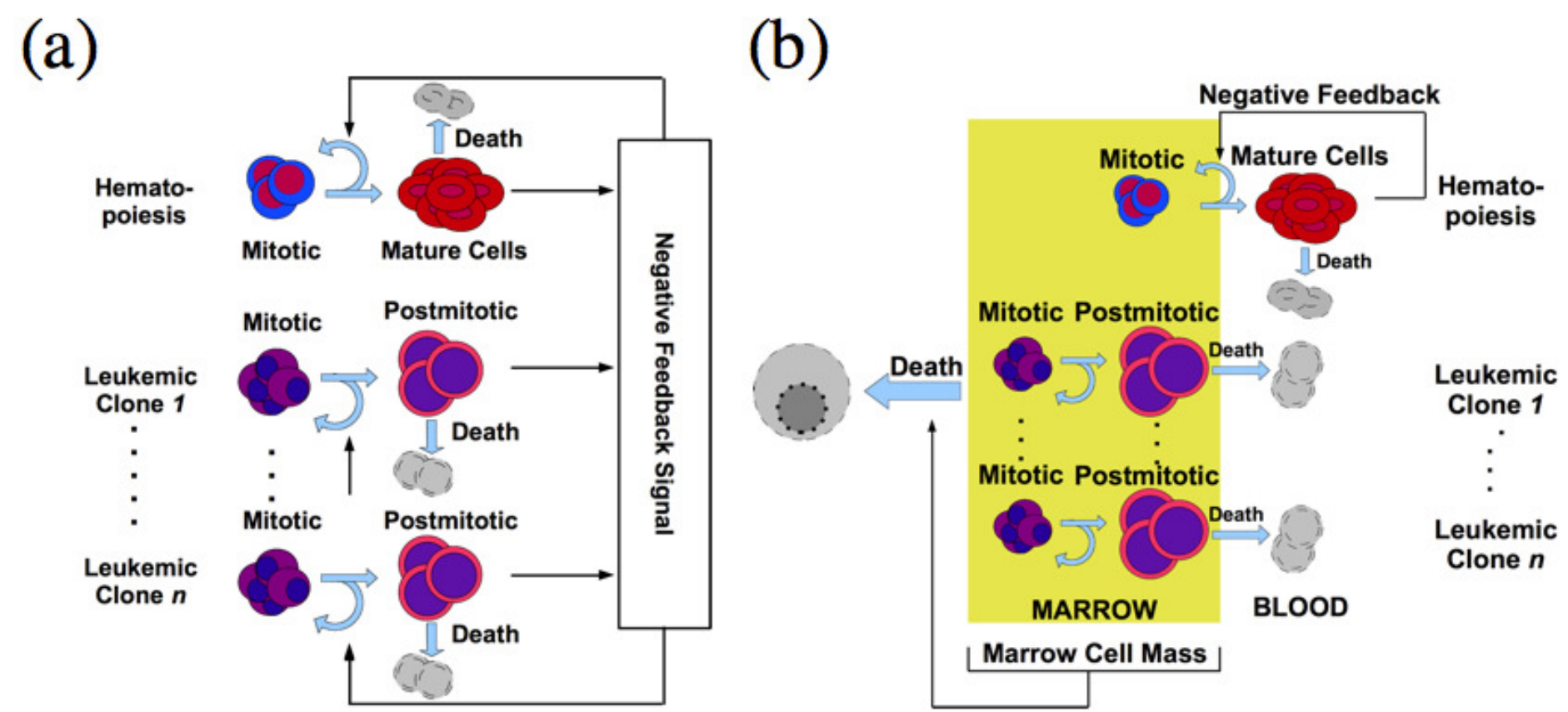} 
\caption{Schematic Representation of the Models: In Model 1 (a) self-renewal of leukemic and hematopoietic cells depends on the total number of postmitotic cells via negative feedback. In Model 2 (b) self-renewal of hematopoietic cells depends on mature  blood cell counts. Leukemic cells are independent of hematopoietic feedback signals. Increasing cell numbers in bone marrow space lead to increasing death rates of all bone marrow cell types, namely mitotic hematopoietic and all leukemic cells. \label{Fig1} }
\end{center}
\end{figure}

\begin{figure}[H] 
\begin{center}
\includegraphics[]{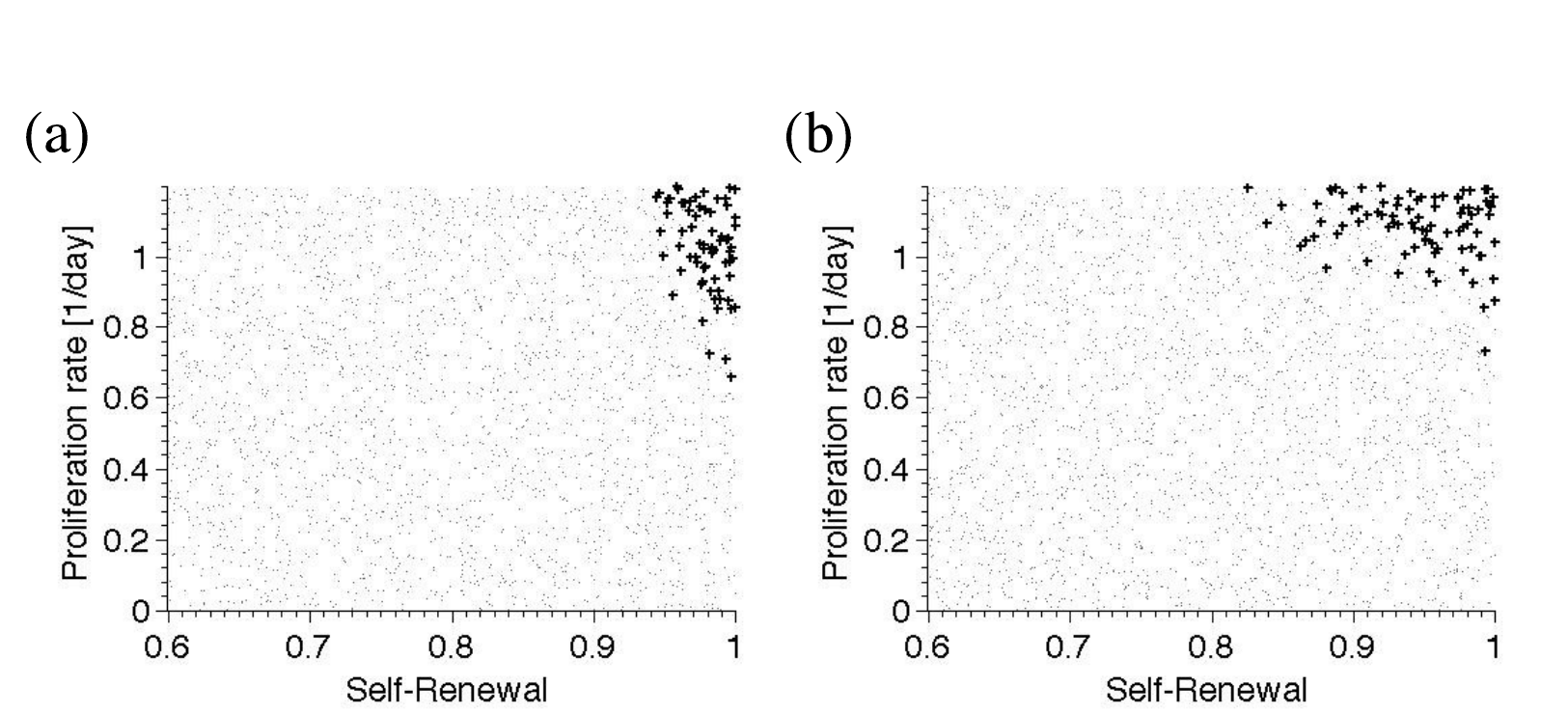} 
\caption{Impact of growth properties on clonal selection.  The figures depict clonal selection in 50 simulated patients. Each black '.' marks cellular properties of a leukemic clone present at the beginning of the simulations in at least one patient. Each '+' marks properties of a leukemic clone contributing significantly to the leukemic cell mass at diagnosis in at least one patient. Leukemic cells present at diagnosis have high proliferation rates and high self-renewal potential. (a): Model 1, (b): Model 2. \label{Fig2}}
\end{center}
\end{figure}

\begin{figure}[H] 
\begin{center}
\includegraphics[]{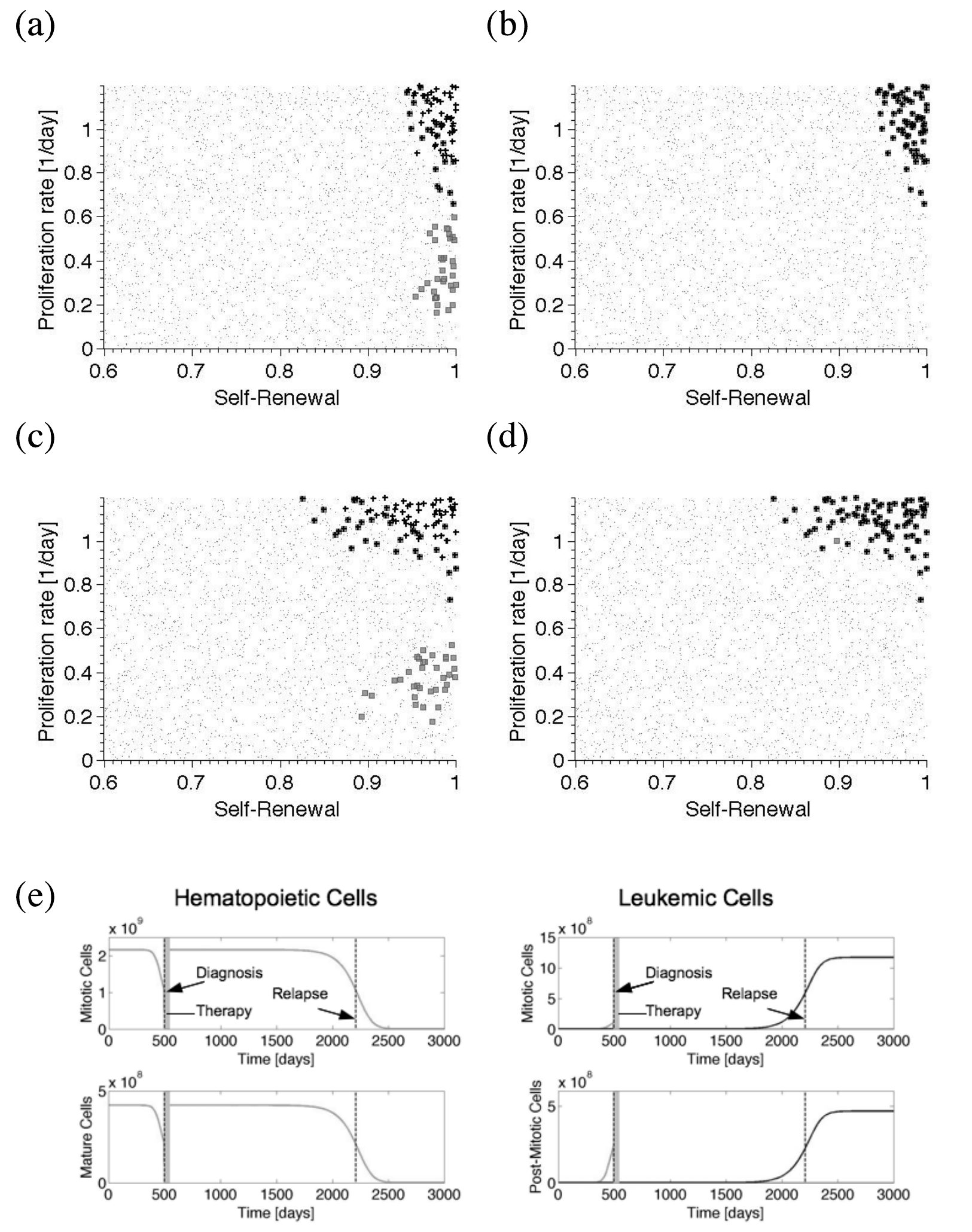} 

\end{center}
\end{figure}

\setcounter{figure}{2}
\begin{figure}[H] 
\begin{center}

\caption{Impact of growth properties on clonal selection. The figures depict clonal selection in 50 simulated patients. Each black '.' marks cellular properties of a leukemic clone present at the beginning of the simulations in at least one patient. Each '+' marks properties of a leukemic clone contributing significantly to the leukemic cell mass at diagnosis in at least one patient. Gray squares mark properties of cell clones contributing significantly to relapse after chemotherapy in at least one patient. In comparison to leukemic cells present at diagnosis, clones at relapse have lower proliferation rates. (a): Model 1, strong chemotherapy, (b): Model 1, weak chemotherapy, (c): Model 2, strong chemotherapy, (d): Model 2, weak chemotherapy. (e) Example of the dynamics of hematopoietic (left) and leukemic (right) cells in one simulated patient. Vertical dotted lines mark primary diagnosis and relapse. Therapy is indicated by a gray rectangle. In the given example primary manifestation and relapse of the disease are diagnosed when mature blood cells decreased by 50\%. \label{Fig3}}
\end{center}
\end{figure}

\begin{figure}[H] 
\begin{center}

\includegraphics[]{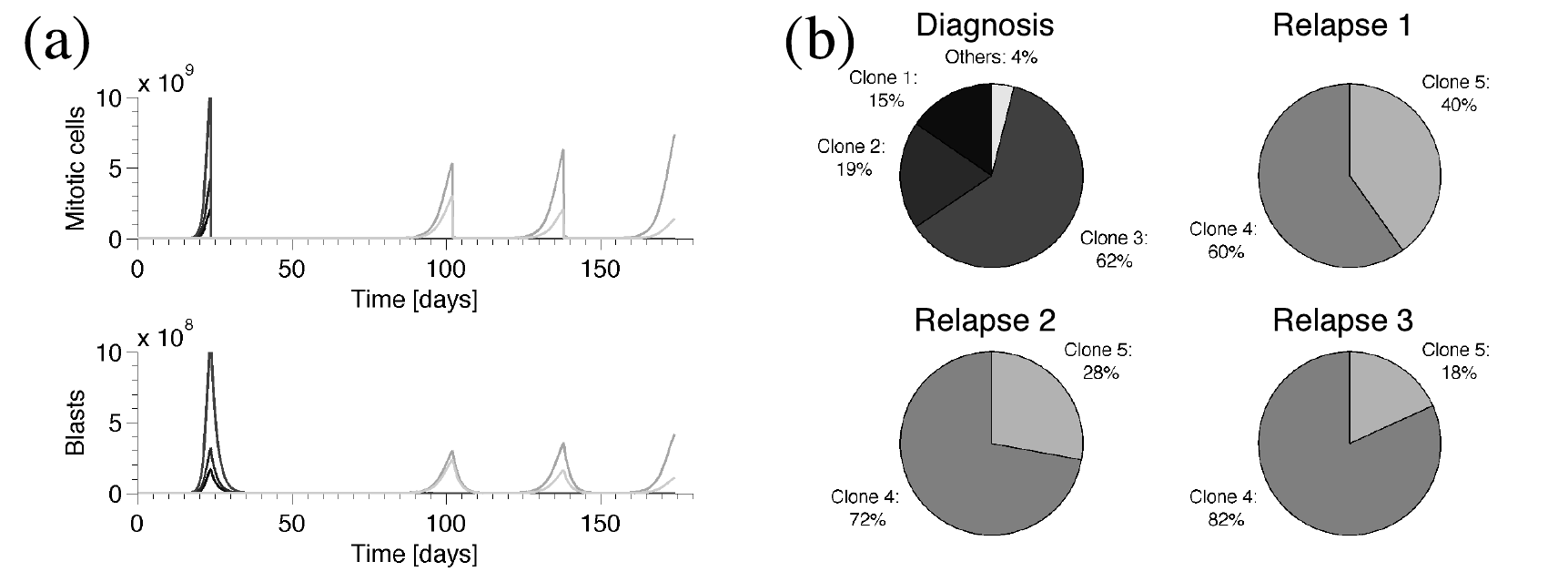} 
\caption{Time dynamics and clonal composition of subsequent relapses. The figure depicts an example of multiple relapses after chemotherapy.  Relapses are treated using the same strategy as primary presentation. (a) Leukemic cell counts, each color indicates a different clone. Time between relapses 2, 3 and 4 is shorter than remission after first treatment. This demonstrates that the selected clones are not fully responsive to the applied therapy. (b) Clonal composition of leukemic cell mass at the primary diagnosis and  at relapses. Charts depict the contribution of major clones to the total leukemic cell mass. Clones responsible for relapse are present at very small fractions at primary diagnosis ($<<5\%$). Relapses are triggered by the same clones but their relative contribution to the leukemic cell mass change in favour of the slowly proliferating highly self-renewing cells. \label{Fig4}}
\end{center}
\end{figure}

\begin{figure}[H]  
\begin{center}
\includegraphics[width=\textwidth]{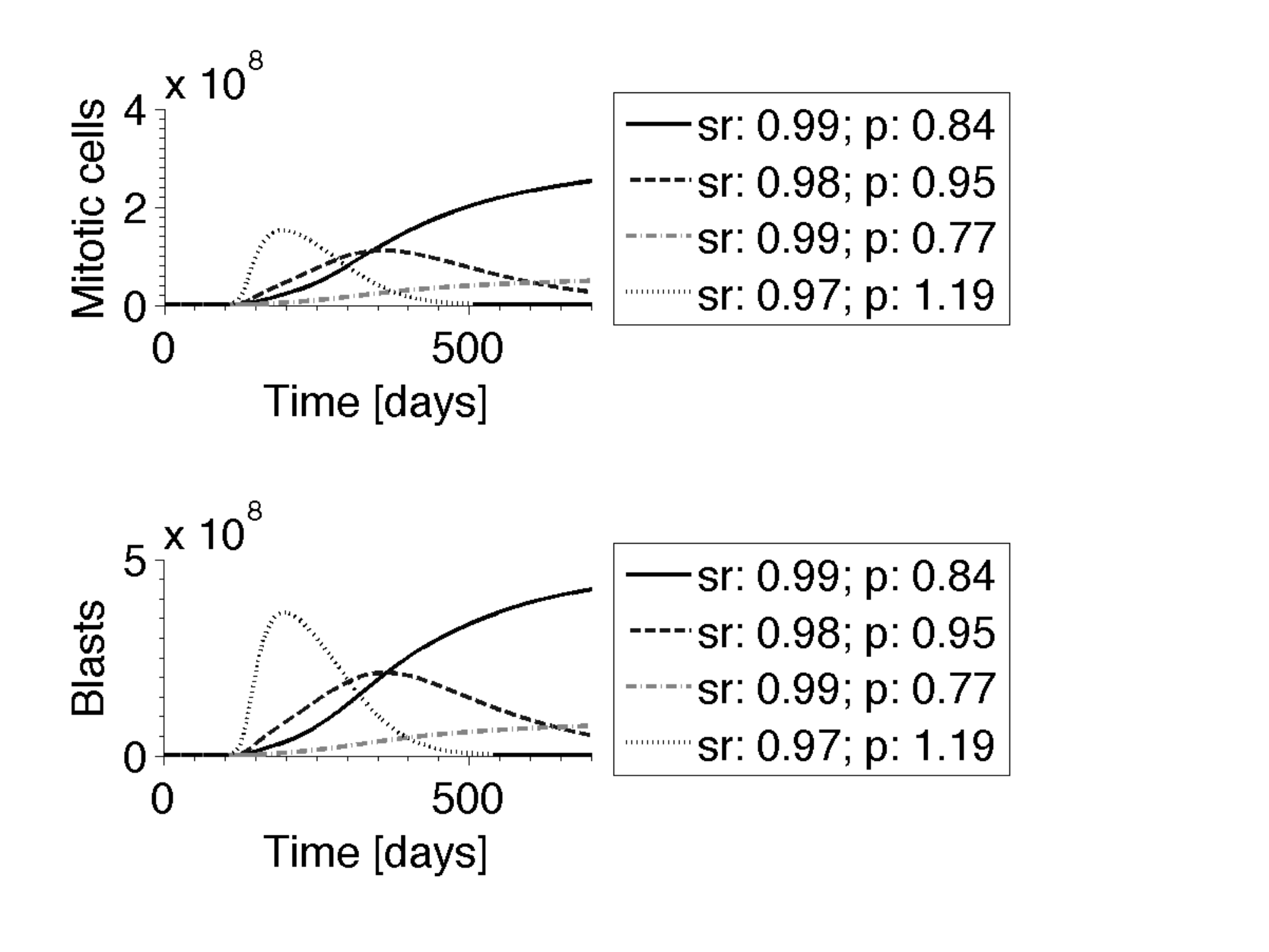} 
\end{center}
\caption{First Phase of Leukemic Clone Evolution: At the beginning fast proliferating clones with low self-renewal can dominate. They are later out-competed by clones with high self-renewal, which is an advantage under high competition for niche spaces, needed for self-renewal. If there exist clones with high self-renewal and high proliferation, they will dominate during this first phase of leukemic evolution. Each line type corresponds to one leukemic clone. Blasts are immature cells used for diagnosis of leukemias. In the course of the disease blasts accumulate and outcompete hematopoiesis. Blast counts greater than $5\%$ are considered as pathological\cite{Liesveld}. The simulations are based on Model 1. \label{Fig5}}
\end{figure}

\pagebreak
\begin{figure}[H]
\includegraphics[]{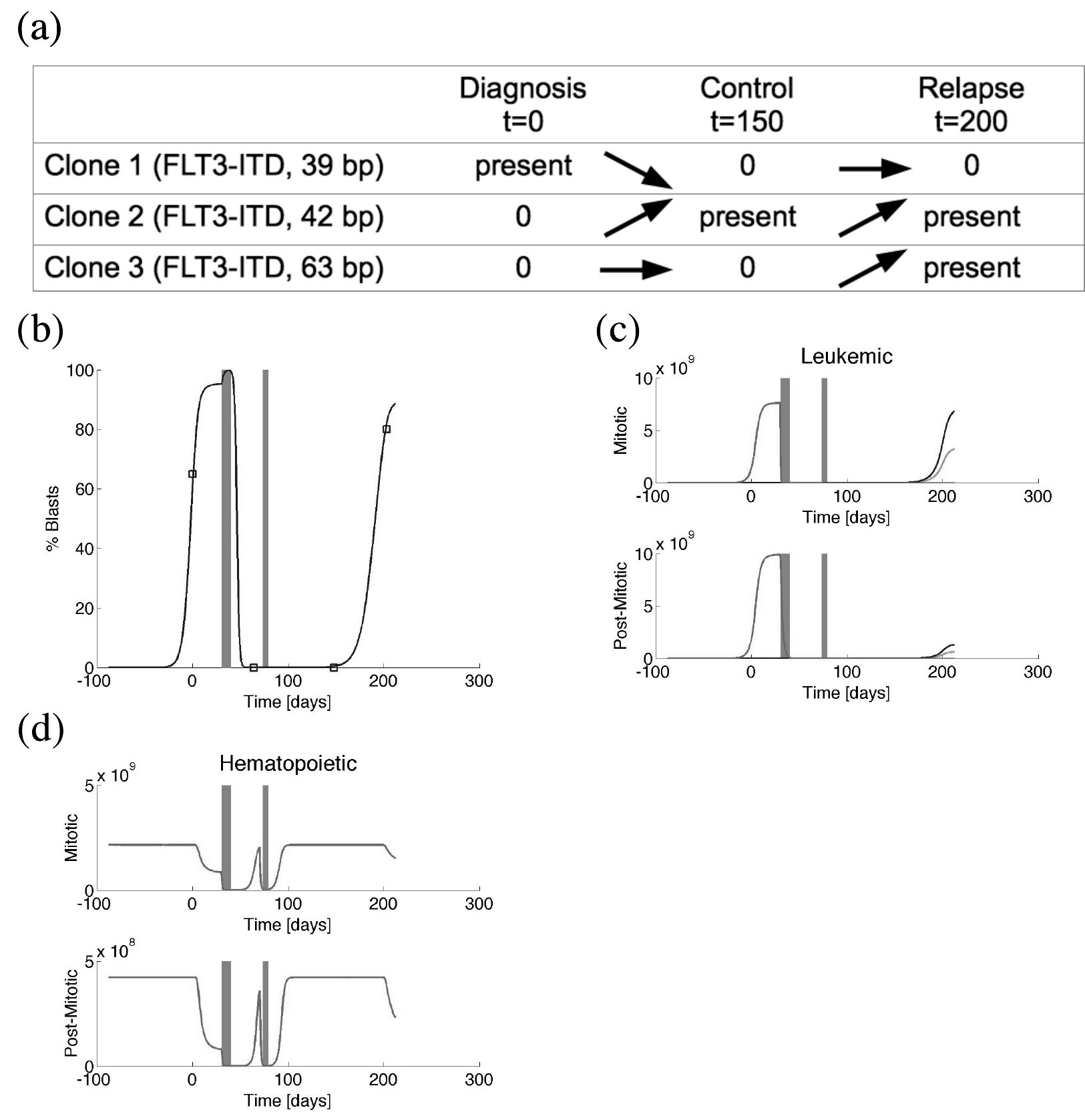} 

\caption{Fitting of model to patient data. Different leukemic mutations are used to distinguish between different clones.  (a)  The table indicates presence and absence of different leukemic clones at different timepoints of the disease. Arrows indicate if the respective clones increased or decreased during the time interval between the measurements. The depicted data are based on PCR analysis of bone marrow cells.  (b) Comparison of simulated blast counts to data. Data are indicated as squares. (c) Evolution of leukemic populations. Each clone is indicated by a different line type. (d) Simulated counts of healthy leukocytes. Chemotherapy cycles are indicated by gray rectangles.}\label{Fig6}
\end{figure}
\thispagestyle{empty} 

\begin{figure}
\includegraphics[]{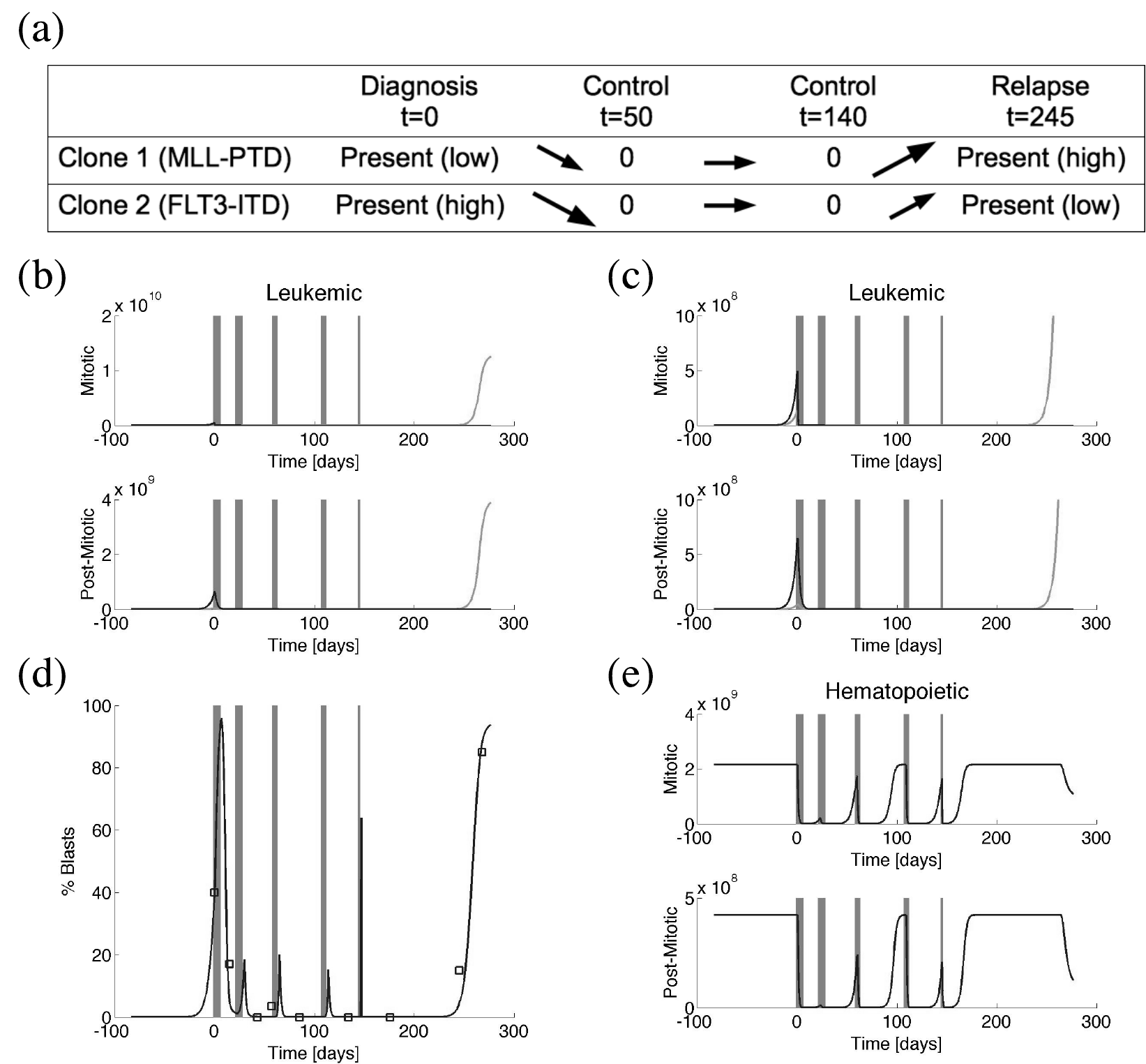} 

\caption{Fitting of model to patient data. Different leukemic mutations are used to distinguish between different clones.  (a) The table indicates presence and absence of different leukemic clones at different timepoints of the disease. Arrows indicate if the respective clones increased or decreased during the time interval between the measurements. Small arrows indicate small changes, large arrows large changes. The depicted data are based on PCR analysis of bone marrow cells. (b)-(c):  Evolution of leukemic populations with differently scaled vertical axis (cells per kg of body weight).  Each clone is indicated by a different line type.
(d): Comparison of simulated blast counts to data. Data are indicated as squares. (e): Simulated counts of healthy leukocytes in cells per kg of body weight. Chemotherapy cycles are indicated by gray rectangles. }\label{Fig7}
\end{figure} 

\setcounter{figure}{0}
\renewcommand{\figurename}{Appendix - Figure}
\begin{figure}
\includegraphics[]{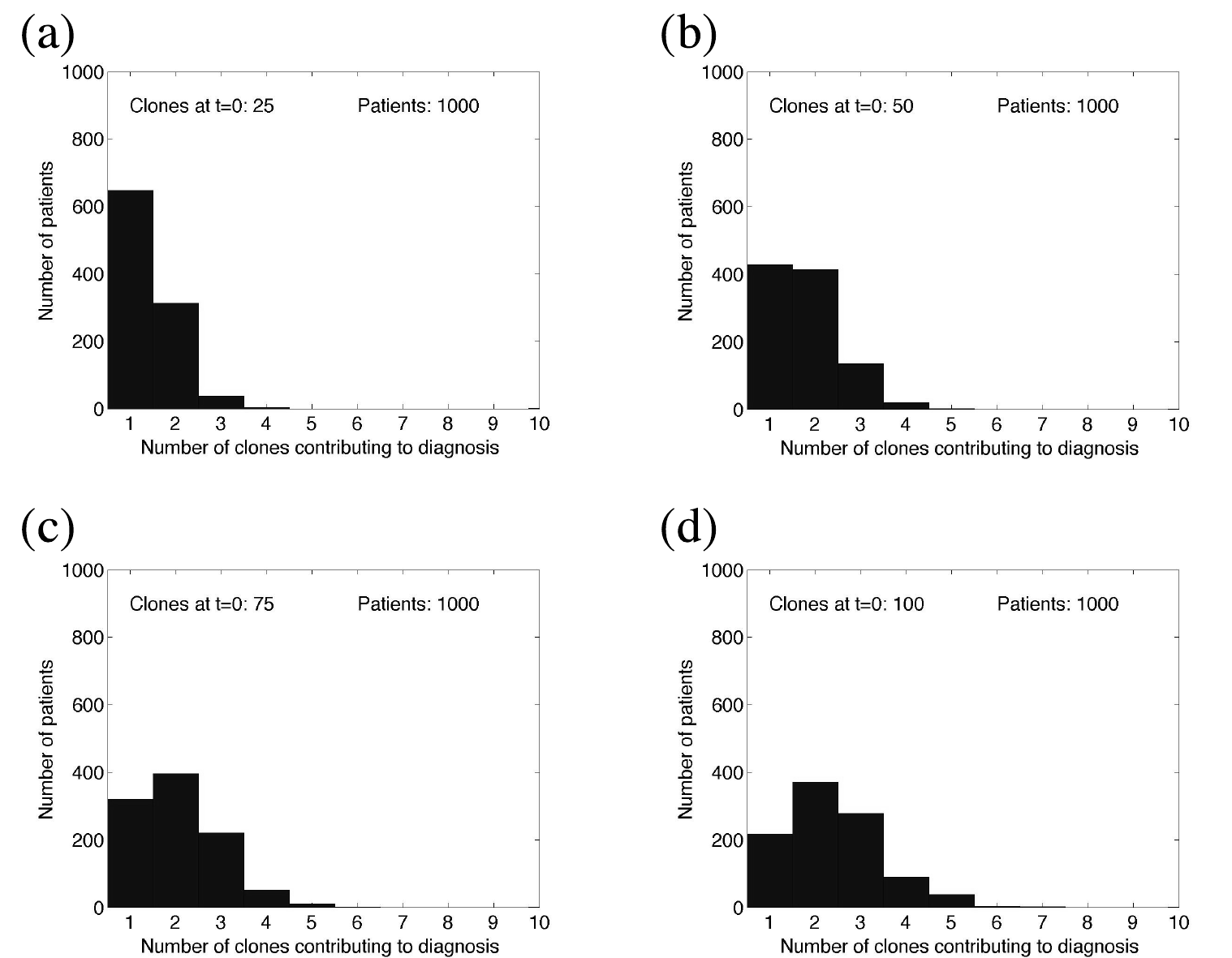} 

\caption{Clones contributing to relapse (Model 1).
The Figure shows the distribution of the number of clones that contribute to diagnosis. The number of clones present at the beginning is 25 (a), 50 (b), 75 (c) or 100 (d). The simulations include 1000 patients. The Figure shows that for all initial conditions (a)-(d) the number of contributing clones is relatively constant (3 or less for more than 80\% of the patients). }\label{SFig1}
\end{figure} 

\begin{figure}
\includegraphics[]{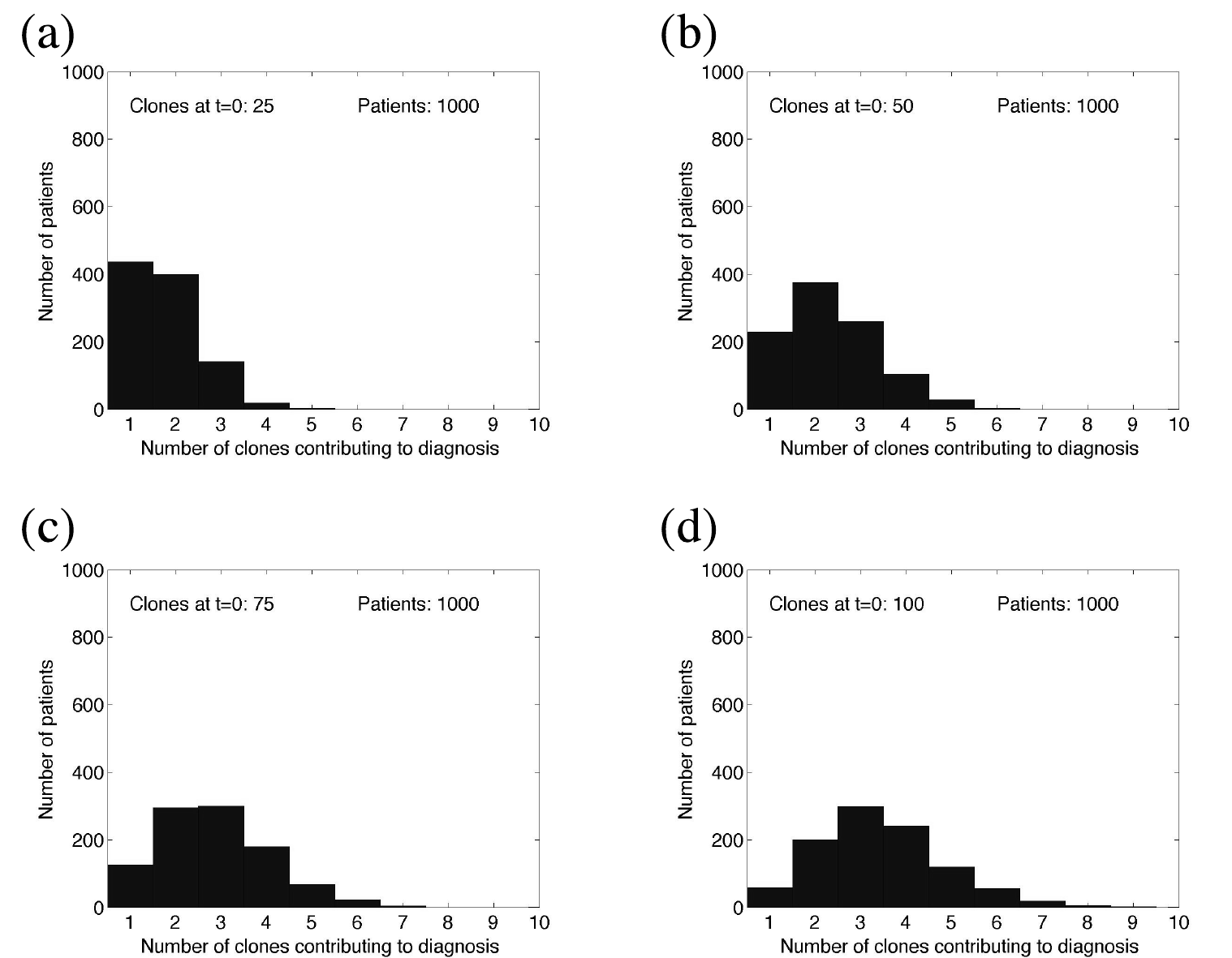} 

\caption{Clones contributing to relapse (Model 2).
The Figure shows the distribution of the number of clones that contribute to diagnosis. The number of clones present at the beginning is  25 (a), 50 (b), 75 (c) or 100 (d). The simulations include 1000 patients. The Figure shows that for all initial conditions (a)-(d) the number of contributing clones is relatively constant (5 or less for more than 80\% of the patients). In Model 2 the average number of clones contributing to relapse is slightly higher than in Model 1.}\label{SFig2}
\end{figure}

\begin{figure}
\includegraphics[]{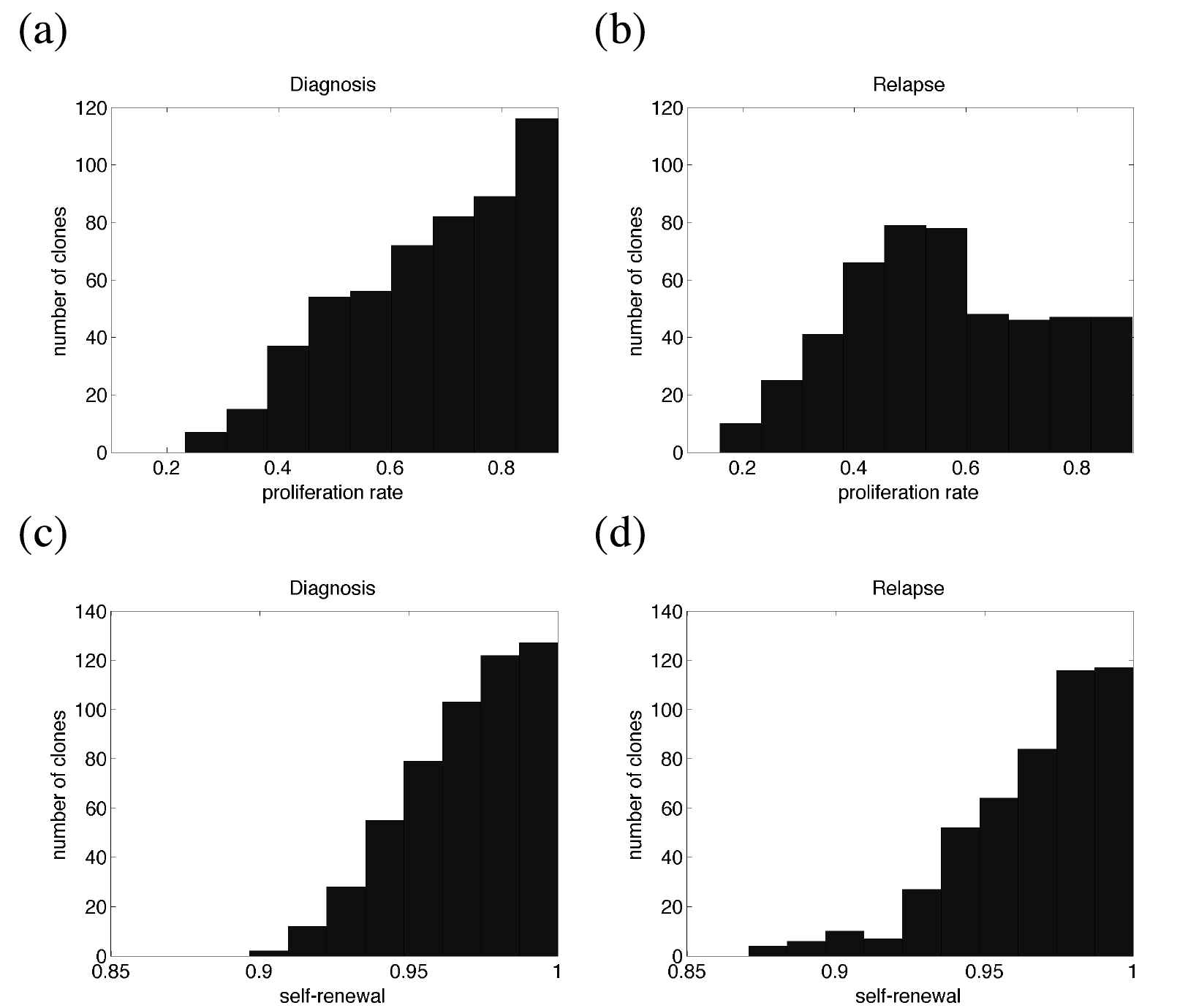} 

\caption{Clonal properties at diagnosis and at relapse in a model with mutations.
The Figure shows the distribution of self-renewal and proliferation rate of leukemic clones present at diagnosis and at relapse. The plots include data of 500 simulated patients. As in the models without mutations, proliferation is reduced at relapse in comparison to diagnosis while self-renewal is high at both timepoints. (a) proliferation rate at diagnosis, (b) proliferation rate at relapse, (c) self-renewal at diagnosis, (d) self-renewal at relapse. The simulations are for $\gamma=0.02/days$ and for $\nu=5\cdot 10^{-8}$, $k_{chemo}=60.$ Similar results are obtained for different values, e.g.,  if $\gamma$ and $\nu$ are varied by a factor of 10. }\label{SFig3}
\end{figure}

\end{document}